\documentclass[10pt,amssymb,superscriptaddress,showkeys,prl, twocolumn, letterpaper]{revtex4-2}

\usepackage{graphicx}
\usepackage{dcolumn}
\usepackage{bm}
\usepackage{amsmath,amssymb}
\usepackage{titlesec}
\usepackage{lineno}
\usepackage{hyperref}
\usepackage{textcomp}
\usepackage{verbatim}

\makeatletter
\renewcommand{\@caption@fignum@sep}{\space} 
\makeatother
\titleformat{\section}[block]{\normalfont\bfseries\large}{\thesection}{1em}{}
\titlespacing*{\section}{0pt}{\baselineskip}{0.5\baselineskip}

\titleformat{\subsection}[block]{\normalfont\bfseries\small}{\thesubsection}{1em}{}
\titlespacing*{\subsection}{0pt}{\baselineskip}{0.5\baselineskip}

\renewcommand{\figurename}{\textbf{Fig.}}  

\hypersetup{
 colorlinks=true,
 linkcolor=blue,
 filecolor=blue,  
 urlcolor=blue,
 citecolor=blue,
}

\begin{document}

\title{Orbital Altermagnetic Photonic Crystal}

\author{Sichang Qiu}
\thanks{These authors contributed equally}
\affiliation{State Key Laboratory of Millimeter Waves, Southeast University, Nanjing, China}

\author{Huichang Li}
\thanks{These authors contributed equally}
\affiliation{State Key Laboratory of Optical Fiber and Cable Manufacture Technology, Department of Electronic and Electrical Engineering, Guangdong Key Laboratory of Integrated Optoelectronics Intellisense, Southern University of Science and Technology, Shenzhen, China}

\author{Yan Meng}
\affiliation{Marine Science and Technology Domain, Beijing Institute of Technology, Zhuhai 519088, China}

\author{Xiang Xi}
\affiliation{School of Electrical Engineering and Intelligentization, Dongguan University of Technology, Dongguan, 523808, China.}

\author{Zebin Zhu}
\affiliation{State Key Laboratory of Optical Fiber and Cable Manufacture Technology, Department of Electronic and Electrical Engineering, Guangdong Key Laboratory of Integrated Optoelectronics Intellisense, Southern University of Science and Technology, Shenzhen, China}

\author{Ce Shang}
\email{shangce@aircas.ac.cn}
\affiliation{Aerospace Information Research Institute, Chinese Academy of Sciences, Beijing 100094, China}

\author{Zhen Gao}
\email{gaoz@sustech.edu.cn}
\affiliation{State Key Laboratory of Optical Fiber and Cable Manufacture Technology, Department of Electronic and Electrical Engineering, Guangdong Key Laboratory of Integrated Optoelectronics Intellisense, Southern University of Science and Technology, Shenzhen, China}

\author{Tie Jun Cui}
\email{tjcui@seu.edu.cn}
\affiliation{State Key Laboratory of Millimeter Waves, Southeast University, Nanjing, China}

\author{Shuo Liu}
\email{liushuo.china@seu.edu.cn}
\affiliation{State Key Laboratory of Millimeter Waves, Southeast University, Nanjing, China}

\date{\today}

\begin{abstract}
Altermagnetism features momentum-dependent spin splitting without net magnetization, extending spintronics beyond conventional ferromagnetism and antiferromagnetism. However, the photonic realization of altermagnetism has remained a formidable challenge due to the fundamental differences between fermionic electrons and bosonic photons. Here, we report the first experimental realization of an orbital altermagnetic photonic crystal, based on an antiunitary $C_{4z}\mathcal{T}$ symmetry enforced correspondence between a local $p$-orbital $\sigma/\pi$ doublet and crystal momentum. We experimentally demonstrate that the resulting system exhibits momentum-dependent spin splitting with alternating pseudospin polarization and a $d_{xy}$-wave form factor, as confirmed by measured band structures and iso-frequency contours. Moreover, we show that the orbital altermagnetic photonic crystal supports unique pseudospin-selective transport of electromagnetic waves, including photonic pseudospin splitting and pseudospin filtering. Our results extend the field of alternagnetism to photonic systems, opening a new avenue for designing \textit{spinphotonic} devices.

\end{abstract}

\maketitle

\section{Introduction}

Altermagnetism has recently emerged as a distinct magnetic phase that goes beyond conventional ferromagnetism and antiferromagnetism \cite{Smejkal2022,Sinova2022,Krempasky2024MnTe,Fedchenko2024RuO2,Amin2024MnTe,LiuNature2026,LeiHan2024,Zhou2025Nature,Gu2025PRL,Jiang2025,Liu2025NatPhy}. Unlike a ferromagnet, which exhibits spin-split bands together with a nonzero net magnetization, and an antiferromagnet, characterized by spin-degenerate bands and a vanishing net magnetization, an altermagnet combines their advantages and supports momentum-dependent spin-split bands despite zero net magnetization. Recent advances have shown that this seemingly paradoxical coexistence is dictated by spin-group symmetry rather than by relativistic spin-orbit coupling, and that the resulting spin splitting generically carries anisotropic $d$-, $g$-, or $i$-wave form factors in momentum space. Thus, altermagnetism is a new symmetry-driven band phenomenon \cite{Jungwirth2026,Smejkal2020CrystalHall,GonzalezHernandez2021SpinSplitter,Cheong2025Classification,Wang2025SpinOrbital}. Because this symmetry-based mechanism combines the stray-field-free advantage of antiferromagnets with the spin-split band functionality of ferromagnets, altermagnetism has rapidly become a central theme in spintronics, magneto-optics, and topological physics  \cite{Bai2022SpinSplittingTorque,Lee2024BrokenKramers}. This naturally raises an interesting question: can such symmetry-enforced momentum-dependent band splitting also be realized beyond electronic systems, particularly in classical wave systems such as photonic crystals?

Photonic crystals provide a promising platform for exploring this possibility. Maxwell’s equations can be formulated as a Hermitian eigenvalue problem, and photonic crystals support Bloch bands, symmetry representations, and momentum-space dispersions that are closely analogous to those of quantum materials \cite{Joannopoulos2008,Raman2010,Khanikaev2013PhotonicTI}. Moreover, the internal degrees of freedom (DoFs) of light waves—including polarization, orbital character, and mode symmetry—are highly tunable in photonic crystals \cite{Bliokh2015SpinOrbitLight,Bliokh2015QSHLight,Dong2017ValleyPC}. Indeed, photonic crystals have been widely adopted to explore various new physics, such as band topology, topologically structured light fields, bound states in the
continuum, non-Hermitian physics, quantum geometry, and photonic time crystals \cite{Khanikaev2013PhotonicTI,LuLing2014-TopologicalPhotonics,Leykam2026-TopologicalPhotonics,Han2024-TopologicalPhotonics,Kang2023-BIC,Miri2019-nonHermitian,Feis2025-SpaceTimephotonic,Lyubarov2022-SpaceTimephotonic}. However, the newly discovered phenomenon of altermagnetism has yet to be experimentally realized in photonic crystal systems.

\begin{figure*}
    \centering
    \includegraphics[width=\linewidth]{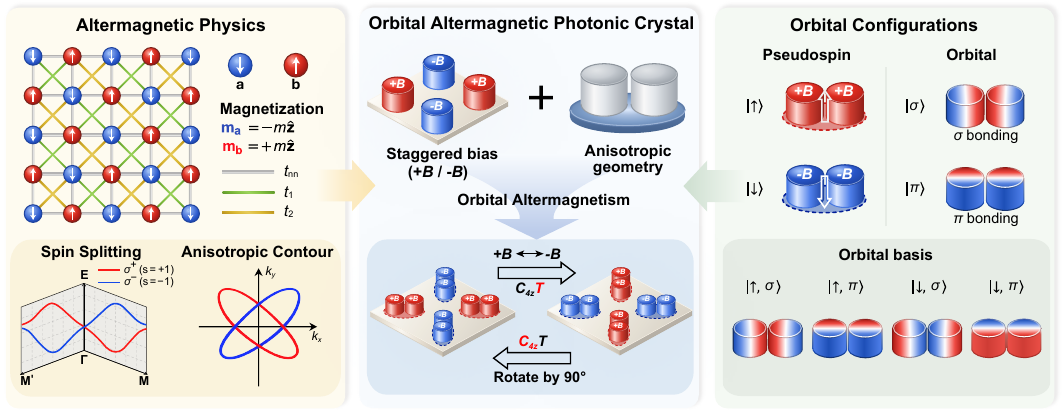}
    \caption{\textbf{Conceptual framework for realizing altermagnetism in photonic crystals}. The left panel shows a minimal altermagnetic tight-binding (TB) model, in which staggered magnetization (blue and red spheres) and anisotropic hopping (green and yellow lines) lead to momentum-dependent band spin splitting (lower-left inset) and anisotropic iso-frequency contours (lower-right inset), capturing the key signature of altermagnetism. The right panel summarizes the orbital configurations of the photonic crystal system: the sign of the magnetic bias (upper-left inset) determines the pseudospin-like component, while the parity of the local mode (upper-right inset) defines the orbital character; together, they form a four-state local basis at each site (lower inset). Based on these ingredients, the central panel presents the orbital altermagnetic photonic crystal: a staggered magnetic bias (red and blue cylinders) introduces a bias-dependent DoF, and replacing isotropic cylindrical elements with anisotropic ones (two grey cylinders) reshapes the local mode profiles and generates the orbital character.}
    \label{fig1}
\end{figure*}

To translate the hallmark band feature of altermagnetism—namely, momentum-dependent spin splitting without net magnetization—into photonic crystals, one must go beyond engineering a generic pseudospin DOF. The pseudospin needs to be defined together with an orbital DOF that provides directional anisotropy, and the two sectors must be organized by a crystal symmetry that enforces alternating pseudospin polarization at symmetry-related momenta. Furthermore, the resulting momentum-space texture should be linked to measurable wave transport in finite photonic crystals, where the excitation condition and boundary orientation determine which Bloch modes are accessed. This demanding scheme motivates a mode-space construction in which pseudospin, orbital character, and crystalline symmetry are designed together as a unified, tightly coupled structure.


Here, we theoretically propose and experimentally demonstrate an orbital altermagnetic photonic crystal, in which the orbital and pseudospin DOFs are both defined in mode space and coupled in a symmetry-controlled manner. Specifically, the pseudospin is defined by the two-component complex amplitudes of the mode on two pairs of identically biased gyromagnetic rods on the same diagonal, analogous to a mode spinor \cite{Song2015,Haldane2008}, whereas the orbital DOF is defined by the local $p$-orbital mode within each pair of identically biased gyromagnetic rods \cite{Gao2023,Zhang2023,Schulz2022}. These two sectors are linked by the antiunitary $C_{4z}\mathcal{T}$ symmetry, which establishes the required correspondence between pseudospin texture, orbital character, and crystal momentum. Moreover, we experimentally observe anisotropic momentum-space spin splitting in the measured band structure and demonstrate novel pseudospin-dependent transport phenomena, including spatial pseudospin splitting and chiral-excitation-controlled pseudospin filtering. This work extends altermagnetic physics to photonics and opens a new route toward \textit{spinphotonic}. 





%

\section{Results} 

To realize orbital altermagnetism in a photonic system, we must map the strict symmetry requirements of electronic altermagnets onto artificial photonic DoFs. In electronic systems, altermagnetic band splitting arises from the interplay of staggered magnetic order, orbital anisotropy, and crystalline symmetry, which collectively produce momentum-dependent spin splitting without net magnetization. The essential requirement is not merely the presence of internal DoFs, but a specific crystal symmetry that forces these DoFs into opposite spin sectors with alternating momentum-space polarization. Achieving this on a photonic platform is highly non-trivial.  While photonic crystals routinely emulate Bloch bands and support engineered DoFs, conventional photonic polarizations and pseudospins cannot capture true altermagnetic physics. A faithful photonic analogue requires that the internal DoFs transform strictly according to the crystal symmetries that connect distinct momentum points.

We establish this rigorous correspondence by engineering a mode space. The local Hilbert space is constructed as a direct product of a pseudospin doublet and an orbital doublet. The pseudospin is defined by the two-component mode amplitudes on dimer channels with opposite magnetic biases ($\uparrow/\downarrow$). To provide the required directional anisotropy, we introduce an orbital DoF governed by the local $p$-orbital bonding character ($\sigma/\pi$) of each dimer (Right panel in Fig.~\ref{fig1}). This yields a composite four-state local basis: $\{|\uparrow,\sigma\rangle, |\uparrow,\pi\rangle, |\downarrow,\sigma\rangle, |\downarrow,\pi\rangle\}$. In this architecture, the staggered magnetic bias acts as the pseudospin, while the anisotropic dimer geometry supplies the orbital anisotropy necessary for an altermagnetic form factor \cite{Smejkal2022,Jungwirth2026,Roig2024,Leeb2024OrbitalOrdering}. To enforce the alternating pseudospin polarization across momentum space, we design the lattice to obey the antiunitary symmetry $C_{4z}\mathcal{T}$—a combined operation of a fourfold rotational and time-reversal symmetry. Because time reversal flips the staggered magnetic bias and the $C_{4z}$ rotation maps $(k_x, k_y)$ to $(k_y, -k_x)$, the invariant lattice is strictly constrained to exhibit momentum-dependent splitting. Consequently, the photonic bands display alternating pseudospin polarization between $C_{4z}\mathcal{T}$-related momenta, successfully capturing both the reciprocal-space texture and the fundamental symmetry mechanism of orbital altermagnetism (Middle panel in Fig.~\ref{fig1}).

To capture the essential band mechanism in a minimal, physically transparent form, we adopt an altermagnetic tight-binding (TB) model that describes the symmetry-allowed coupling between the pseudospin and orbital sectors (Left panel in Fig. \ref{fig1}). The corresponding \(d_{xy}\)-wave type altermagnetic Hamiltonian is written as
\begin{equation}
H(k)=\epsilon_0(k)s_0\otimes\tau_0+d_z(k)s_0\otimes\tau_z+\gamma(k)s_0\otimes\tau_x-\Delta s_z\otimes\tau_z.
\label{eq:tb}
\end{equation}
Here, \(s_i\) and \(\tau_i\) denote the Pauli matrices acting on the pseudospin and orbital sectors, respectively. The momentum-dependent terms are given by
\begin{align}
\epsilon_0(k) &= 2(t_1+t_2)\cos k_x \cos k_y, \\
d_z(k) &= -2(t_1-t_2)\sin k_x \sin k_y.
\end{align}
and 
\begin{equation}
    \gamma(k)=2t_{\mathrm{nn}}(\cos k_x+\cos k_y),
\end{equation}
where \(t_1\) and \(t_2\) describe anisotropic diagonal couplings, \(t_{\mathrm{nn}}\) denotes nearest-neighbor hopping, and \(\Delta\) represents the effective splitting induced by the staggered magnetic bias. The $d_z(k)$ term follows a $d_{xy}$-type angular form and therefore produces the characteristic angle-dependent band splitting. The bias-induced term \(\Delta s_z\otimes\tau_z\) lifts the pseudospin degeneracy in a symmetry-sensitive manner. Together, they produce momentum-dependent splitting with alternating pseudospin polarization while preserving zero net magnetization, which is the defining band signature of altermagnetism.

\begin{figure*}
    \centering
    \includegraphics[width=\linewidth]{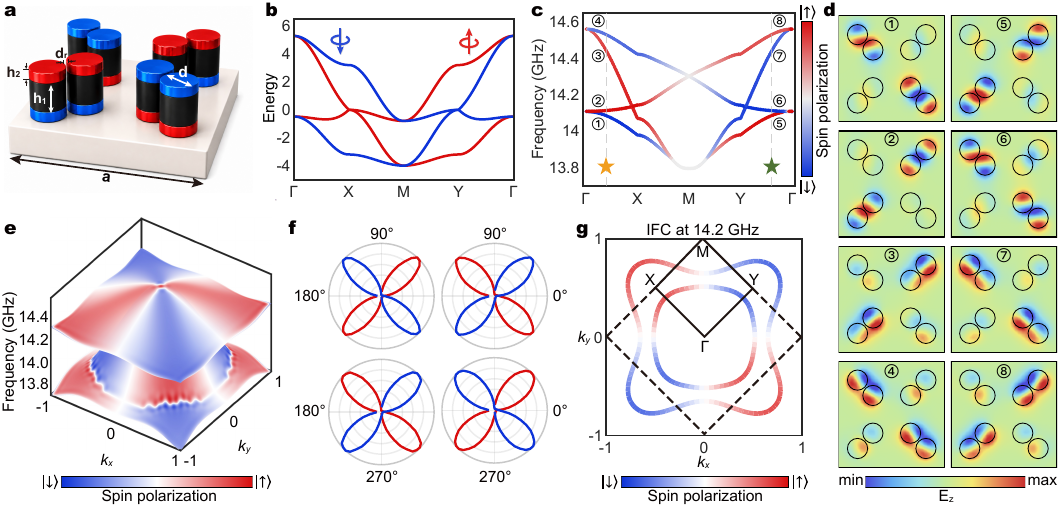}
    \caption{\textbf{Design of an orbital altermagnetic photonic crystal}. \textbf{a} Unit cell of the altermagnetic photonic crystal featuring anisotropic geometry and staggered magnetic bias with a lattice constant of $a=21\,$mm. Each site consists of a YIG cylinder (diameter $d=4\,$mm, height $h_1=3\,$mm, separated by an intra-site distance $d_r$) sandwiched between two magnets (diameter $d=4\,$mm, height $h_1=1\,$mm). \textbf{b} Calculated TB bulk band structure based on Eq.~(\ref{eq:tb}) with $t_1=0.2,\,t_2=1,\,t_{nn}=0.6,\,\Delta=2(t_2-t_1)$. The spin-resolved band structure exhibits characteristic momentum-dependent band splitting along high-symmetry paths. \textbf{c} Simulated bulk band structures of the orbital altermagnetic photonic crystal, reproducing the key features of the TB model. The highlighted stars mark a pair of $C_{4z}\mathcal{T}$-related representative momenta: $\mathbf{k}_1=(-\pi/5,\pi/5)$ along the $\Gamma-X$ path (yellow star) and $\mathbf{k}_2=(\pi/5,\pi/5)$ along $\Gamma-Y$ path (green star). \textbf{d} Simulated eigenmode profiles $E_z$ at eight frequencies corresponding to the highlighted momenta in \textbf{c}. The upper (lower) panels correspond to $\mathbf{k_1}$ ($\mathbf{k_2}$). \textbf{e} Three-dimensional dispersion surface with spin-resolved projections, revealing the momentum-dependent pseudospin polarization characteristic of altermagnetic band splitting. \textbf{f} Angular dependence of the spin projection for the four bands evaluated along a constant-$|\mathbf{k}|$ contour with $|\mathbf{k}|= 0.6$ from the altermagnetic photonic crystal dispersion, showing a pronounced anisotropic four-lobed pattern. \textbf{g} Iso-frequency contour (IFC) at $f$=14.2~GHz, highlighting the anisotropic spin-momentum texture.}
    \label{fig2}
\end{figure*}

\subsection{Design of an orbital altermagnetic photonic crystal}

To realize the required symmetry correspondence, we design a photonic crystal unit cell that jointly encodes pseudospin and orbital DoFs in mode space. The structure combines staggered magnetic bias with anisotropic resonator geometry, as illustrated in Fig.~\ref{fig2}a. It consists of four resonant elements arranged on a square lattice with lattice constant \(a\). Each element is formed by a pair of yttrium iron garnet (YIG) cylinders sandwiched by permanent magnets, producing a gyromagnetic dimer resonator \cite{Wang2008,Xi2023,Zhou2024,Wang2009OneWayObservation,Skirlo2015LargeChern}. The magnetic bias alternates between neighboring dimers, thereby defining two dimer channels with opposite bias.

The TB band structure obtained from Eq.~(\ref{eq:tb}) is shown in Fig.~\ref{fig2}b. It exhibits a clear momentum-dependent splitting between the two pseudospin sectors along high-symmetry directions. The corresponding pseudospin polarization alternates across the BZ. This behavior originates from the \(d_{xy}\)-type anisotropy and the symmetry constraint imposed by \(C_{4z}\mathcal{T}\), under which symmetry-related momentum points carry opposite pseudospin polarization while remaining degenerate in frequency. The full-wave simulated dispersion in Fig.~\ref{fig2}c reproduces the essential features predicted by the TB model. In particular, it exhibits a clear momentum-dependent band splitting together with alternating pseudospin polarization across the BZ. To further clarify the underlying physics, we examine two representative momentum points, \(k_1=(-\pi/5,\pi/5)\) and \(k_2=(\pi/5,\pi/5)\), highlighted in Fig.~\ref{fig2}c. These two points are connected by the antiunitary \(C_{4z}\mathcal{T}\) symmetry and therefore satisfy $\omega(\mathbf{k})=\omega(C_{4z}\mathcal{T}\mathbf{k})$. At the same time, they carry opposite pseudospin polarization, consistent with the symmetry-constrained band response expected for photonic orbital altermagnetism.

The simulated $E_z$ profiles of the Bloch eigenmodes in Fig.~\ref{fig2}d provide a real-space view of DOSs, identified from the dimer channel characterized by localized modes. The orbital DoF is identified from the field distribution inside the localized dimer channel. In each dimer, the relative field distribution between the two cylinders forms a characteristic bonding pattern, which can be associated with the local $p$-orbital $\sigma$- or $\pi$-type modes introduced in Fig.~\ref{fig1}. The $C_{4z}\mathcal{T}$ symmetry relates opposite pseudospin sectors at symmetry-connected momenta, whereas the local dimer geometry keeps the orbital character well defined.

\begin{figure*}
    \centering
    \includegraphics[width=\linewidth]{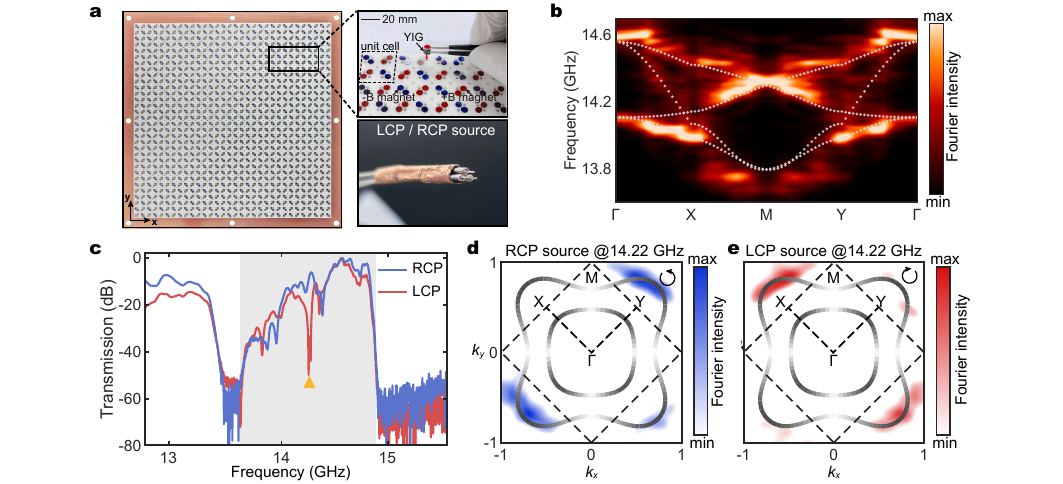}
    \caption{\textbf{Experimental realization of an orbital altermagnetic photonic crystal}. \textbf{a} Left panel: photograph of the fabricated sample, where each YIG cylinder is biased by a pair of permanent magnets. Opposite magnetization directions (red and blue dots) define the two pseudospin sectors. The dashed square marks one unit cell. Right panels: close-up view of a magnet-biased YIG rod and the chiral source used to generate RCP- and LCP excitations. \textbf{b} Measured (colormap) and simulated (grey dots) bulk band structure of the orbital altermagnetic photonic crystal along the high-symmetry path $\Gamma-X-M-Y-\Gamma$. \textbf{c} Measured transmission spectra $S_{21}$ at the same probe position for RCP (blue line) and LCP (red line) excitations. Near 14.2~GHz (yellow triangle), a clear signal is detected for RCP excitation, while the LCP signal is strongly suppressed. This contrast indicates that only the RCP-excited pseudospin component efficiently reaches this probe position. The shaded region highlights the frequency window used for the field and Fourier analyses. \textbf{d,e} Iso-frequency contours at 14.22~GHz extracted from the Fourier spectra for RCP and LCP excitations coupled to opposite-bias YIG sites. The two sources selectively excite opposite pseudospin components, revealing their distinct momentum-space distributions on the same iso-frequency contour.}
    \label{fig3}
\end{figure*}

The global momentum-space signature of this mechanism is shown in Figs.~\ref{fig2}e--g. The pseudospin-resolved dispersion reveals a continuous evolution of pseudospin polarization across the BZ, with opposite polarization appearing at $C_{4z}\mathcal{T}$-related momenta. The extracted angular dependence exhibits a pronounced fourfold pattern, consistent with the \(d_{xy}\)-wave form factor (see Supplementary Materials for details). The corresponding isofrequency contours are strongly anisotropic, providing a direct visualization of the altermagnetic pseudospin texture in the photonic crystal.

The TB analysis and full-wave simulations reveal that the observed band response is not solely due to structural anisotropy. It is a symmetry-governed, momentum-dependent splitting with alternating pseudospin polarization, arising from the combined action of the dimer-channel pseudospin, the local $p$-orbital $\sigma/\pi$ anisotropy, the staggered magnetic-bias pattern, and the antiunitary $C_{4z}\mathcal{T}$ symmetry.

\begin{figure*}
    \includegraphics[width=\linewidth]{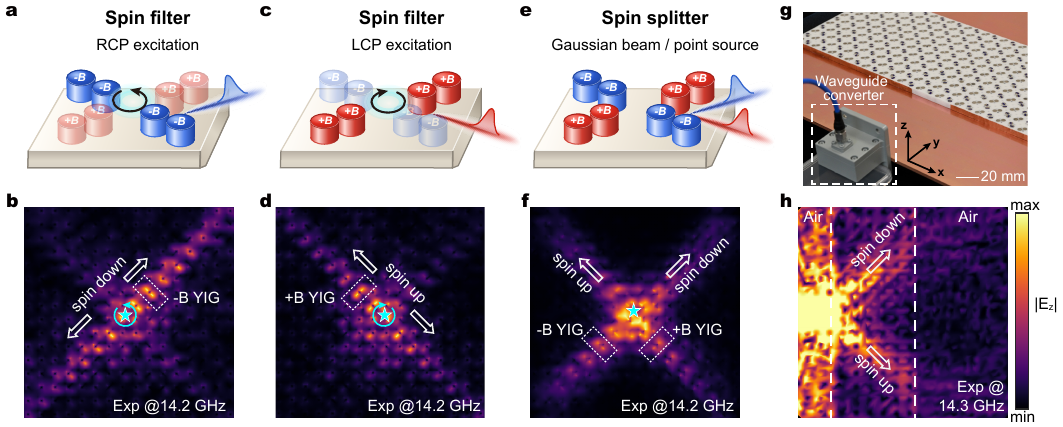}
    \caption{\textbf{Observation of spin-splitting and spin-filtering in an orbital altermagnetic photonic crystal}. \textbf{a} Schematic of spin filtering under RCP excitation. The chiral RCP excitation selectively couples to the spin-down channel, giving rise to directional transport with the field predominantly localized near the $-B$-biased YIG sites. \textbf{b} Measured $|E_z|$ field distribution for the RCP excitation configuration at 14.2~GHz. The white arrows indicate the spin-down propagation channel, and the dashed box marks representative $-B$-biased YIG sites. \textbf{c} Schematic of spin filtering under LCP excitation. Reversing the excitation handedness selects the opposite spin channel, producing opposite-directional transport with the field predominantly localized near the $+B$-biased YIG sites. \textbf{d}  Measured $|E_z|$ field distribution for the LCP excitation configuration at 14.2~GHz. The white arrows indicate the spin-up channel, and the dashed box marks representative $+B$ YIG sites. \textbf{e} Schematic of the spin-splitter under nonchiral excitation, including point-source and Gaussian-beam configurations. The excitation couples to both spin channels, which separate into two diagonal propagation paths. \textbf{f} Measured $|E_z|$ field distribution under point-source excitation at 14.2~GHz. The white arrows indicate the spatial separation of the spin-up and spin-down channels, which are predominantly localized near the $+B$- and $-B$-biased YIG sites, respectively. \textbf{g} Experimental configuration for spin-splitting measurements under Gaussian-beam excitation. A waveguide converter launches the incident microwave beam into the finite photonic crystal sample. \textbf{h}  Measured $|E_z|$ field distribution under Gaussian-beam excitation at 14.3~GHz, showing spin splitting of the transmitted field into two separated propagation channels.}
    \label{fig4}
\end{figure*}

\subsection{Experimental realization of an orbital altermagnetic photonic crystal}

To verify the altermagnetic band structure in a finite photonic crystal, we performed near-field measurements on a fabricated $16\times16$ array using the experimental setup shown in Fig.~\ref{fig3}a. The main image shows a portion of the fabricated sample. The insets show a magnet-biased YIG element and the chiral source used to generate RCP and LCP excitations. 

Figure~\ref{fig3}b shows the Fourier-reconstructed dispersion along the high-symmetry path \cite{Engelen2005BlochDispersion,Abashin2006NearFieldModes,Wu2017ValleyPolarized}. The measured Fourier intensities exhibit clear band-like features that agree well with the simulated bulk dispersion, as indicated by the white dotted lines. This agreement confirms that the finite array preserves the essential Bloch-band response of the photonic crystal, including the anisotropic momentum dependence associated with altermagnetic band splitting.

We then measured the transmission spectra at the same probe position under RCP and LCP excitations, as shown in Fig.~\ref{fig3}c. Near 14.2~GHz, the RCP signal is clearly transmitted, whereas the LCP signal is strongly suppressed, indicating pseudospin-selective propagation.

To connect the real-space propagation with the momentum-space band structure, we extracted the isofrequency contours at $f=14.22$~GHz from the Fourier spectra, as shown in Figs.~\ref{fig3}d and \ref{fig3}e. The RCP and LCP sources preferentially excite opposite pseudospin components on the same iso-frequency contour, resulting in distinct momentum-space intensity patterns. The contours display a pronounced fourfold anisotropy, consistent with the $d_{xy}$-type form factor of the altermagnetic splitting. The anisotropic iso-frequency contours indicate that the excited modes have different dominant group-velocity directions, which is consistent with the pseudospin-dependent beam propagation observed in real space.

Together, these measurements establish a direct correspondence between finite-array wave propagation and the symmetry-governed altermagnetic band structure. The Fourier-reconstructed dispersion reproduces the bulk band features, while the iso-frequency contours reveal pseudospin-selective occupation and fourfold anisotropy. These results confirm that the finite photonic crystal retains the key momentum-space signatures of photonic orbital altermagnetism and converts them into measurable pseudospin-dependent transport.

\subsection{Observation of altermagnetic spinphotonics functionalities}

The altermagnetic photonic crystal converts its anisotropic pseudospin texture in momentum space into spin-dependent wave transport in real space. As shown in Fig.~\ref{fig4}, we experimentally demonstrate two representative functionalities: chiral-excitation-induced spin filtering and spin splitting under nonchiral excitation. These effects originate from the same pseudospin-dependent iso-frequency structure, but they probe different aspects of the momentum-space response. Chiral excitation selectively addresses one pseudospin channel, whereas nonchiral excitation couples to both pseudospin channels and reveals their spatial separation.~\cite{Dong2017ValleyPC,Chen2017ValleyContrasting,Kang2018PseudoSpinValley,Noh2018ValleyHall}.

We first demonstrate spin filtering using chiral RCP/LCP excitations. Figure~\ref{fig4}a schematically shows the RCP excitation configuration. The RCP source preferentially excites the spin-down channel, leading to directional transport with the $|E_z|$ field predominantly localized near the $-B$-biased YIG sites (Fig.~\ref{fig4}b). Reversing the chirality of the excitation preferentially excites the spin-up channel, producing directional transport in the opposite channel with the field predominantly localized near the $+B$-biased YIG sites (Fig.~\ref{fig4}c). The measured $|E_z|$ distribution in Fig.~\ref{fig4}d confirms the switched response, with the field mainly localized around the $+B$-biased YIG sites.

We then demonstrate spin splitting under nonchiral excitation. Figure~\ref{fig4}e illustrates the spin-splitting mechanism under Gaussian-beam or point-source excitation. Unlike the chiral RCP/LCP sources, a nonchiral excitation couples to both pseudospin channels. Owing to the altermagnetic spin splitting, the two pseudospin channels acquire different group-velocity directions and therefore separate into two diagonal paths. Figure~\ref{fig4}f shows the measured $|E_z|$ field distribution under point-source excitation at 14.2~GHz. The field separates into two spatially distinct branches, as indicated by the white arrows. The two branches are predominantly localized around the $+B$- and $-B$-biased YIG sites, respectively, confirming their opposite pseudospin characters. To further verify this splitting under beam excitation, we use the experimental configuration shown in Fig.~\ref{fig4}g, where a waveguide converter launches an incident microwave beam into the photonic crystal. The measured field distribution at 14.3~GHz (Fig.~\ref{fig4}h) shows that the transmitted Gaussian beam splits into two spin-dependent propagation channels.

\section{Conclusion}



In conclusion, we have experimentally demonstrated the first orbital altermagnetic photonic crystal, extending the concept of altermagnetism from fermions (electrons) to bosons (photons). In this system, the two dimer channels define a pseudospin DoF, while the local $\sigma$- and $\pi$-orbital modes provide an additional orbital DoF. Their symmetry-constrained coupling gives rise to momentum-dependent spin splitting. This coupling produces an orbital altermagnetic pseudospin texture without requiring intrinsic electronic spin or net magnetization. We further demonstrate pseudospin-dependent transport, including spin filtering under chiral excitation and spin splitting under nonchiral excitation. These results show that altermagnetic functionality can be achieved in a photonic system through symmetry and mode engineering. We envision that this orbital altermagnetic photonic crystal can be extended to terahertz and optical frequencies. More broadly, this work provides a symmetry-based route to internal-state control in photonic crystals. The underlying mechanism relies on symmetry engineering in mode space rather than on material-specific electronic interactions. This feature makes the approach readily extendable to other classical-wave systems, such as acoustics and mechanics. Hence, our work not only establishes a new platform to explore altmagnetic physics but also opens a new route to the design of \textit{spinphotonic} devices for light-wave manipulation.

\section*{Methods}
\subsection*{Numerical simulation}

All numerical results presented in this work are simulated using the RF module of COMSOL Multiphysics. The bulk band structure is simulated using a square unit cell with periodic boundary conditions in all directions. Both the copper plates and permanent magnets are modeled as PEC. Under a static magnetic bias along the $z$ direction, the relative permeability tensor of the YIG ferrite was written as:

$$\boldsymbol{\mu}_{\pm}
=
\begin{pmatrix}
\mu_r & \pm i\kappa & 0 \\
\mp i\kappa & \mu_r & 0 \\
0 & 0 & 1
\end{pmatrix},$$
where the sign of the off-diagonal term is determined by the bias direction. The tensor elements were
$$
\mu_r
=
1+
\frac{(\omega_0+i\alpha\omega)\omega_m}
{(\omega_0+i\alpha\omega)^2-\omega^2},$$
and 
$$\kappa
=
\frac{\omega\omega_m}
{(\omega_0+i\alpha\omega)^2-\omega^2},$$
with $\omega_m=\gamma\mu_0M_s$, $\omega_0=\gamma\mu_0H_z$ and $\mu_0H_z$ is the external magnetic field (0.16 Tesla) along the $z$ direction, $\gamma=1.759\times10^{11}~\mathrm{C}/kg $ is the gyromagnetic ratio, $\alpha=0.0088$ is the damping coefficient, and $\omega$ is the operating frequency. 


\subsection*{Pseudospin and angular analysis}
For simulated eigenmodes, the pseudopin polarization of the $n$-th band was calculated as $$S_n(\mathbf{k})
=
\frac{
I_+^{(n)}(\mathbf{k})-I_-^{(n)}(\mathbf{k})
}{
I_+^{(n)}(\mathbf{k})+I_-^{(n)}(\mathbf{k})
},$$
where $I^{(n)}_+$ and $I_-^{(n)}$ are the field weights on the two opposite-bias dimer channels. The angular dependence was obtained by averaging $S_n(k)$ within angular bins on a thin momentum shell around a fixed radius $k_0$. The radial coordinate in the polar plot was nomalized by the maximum absolute value of the angularly averaged pseudospin polarization.

\subsection*{Materials and experimental setups}

In the experiment, we adopt commercially available gyromagnetic materials (YIG ferrites) and permanent magnets to break the time-reversal symmetry. The radius and height of YIG ferrites are 2 mm and 3 mm, respectively. The YIG ferrites have a saturation magnetization Ms = 1780 Gauss. Their relative permittivity ($\varepsilon_{\rm YIG}=14.3+0.003i$) and permeability ($\mu_r \approx 1$) remain nearly constant across microwave frequencies. The permanent magnets ($\mathrm{Sm}_{2}\mathrm{Co}_{17}$) are electroplated by nickel with a thickness of 0.002 mm. One pair of magnets provide an overall uniform external magnetic field of about 0.16 Tesla to magnetize the YIG rods. The copper plates are fabricated by depositing a 0.035 mm-thick copper layer onto a Teflon woven-glass fabric laminate substrate and perforating the plates with air holes using a laser-cutting technique. To fix the positions of the sandwiched YIG rods and permanent magnets, we adopt the perforated dielectric foam (ROHACELL 31 HF) with relative permittivity 1.04 and loss tangent 0.0025. 

In the experimental measurements, microwave dipole antennas functioning as source and probe are connected to a vector network analyzer (Keysight E5080). By inserting the probe into the air holes one by one and scanning along the $x$-$y$ plane in small steps, we can map the complex electric field distributions within the bulk of the experimental sample. For chiral excitation measurements, the source, depicted in Fig.~\ref{fig3}a, comprises three dipole antennas tightly bundled together with a precisely controlled mutual phase difference of $2\pi/3$, which generates a well-defined circularly polarized chiral field. To obtain the projected bulk structures, 2D Fourier transformation was performed on the measured complex electric field distributions at each frequency. For the device characterization, a waveguide converter was employed to transform the emission from the microwave source antenna into a Gaussian wave that illuminates the device.

\section*{Acknowledgements}

S.Q., S.L., and T.J.C. acknowledge support from the National Key Research and Development Program of China (Grant Nos. 2022YFA1404903 and 2023YFB3811504) and the National Natural Science Foundation of China (Grant No. U22A2001 and Grant No. 62288101). Z.G. acknowledges support from the National Key Research and Development Program of China (Grant No. 2025YFA1412300), the National Natural Science Foundation of China (Grant Nos. 62361166627 and 62375118), the Guangdong Basic and Applied Basic Research Foundation (Grant No. 2024A1515012770), the Shenzhen Science and Technology Innovation Commission (Grant No. 20230802205352003), and the High-level Special Funds (Grant No. G03034K004). C.S. acknowledges projects of the Chinese Academy of Sciences (No. E4BA270100, E4Z127010F, E4Z6270100, E53327020D).

\section*{Data availability statement}
The datasets generated and analyzed in the current study are available from the corresponding author upon reasonable request.

\section*{Competing interests}
The authors declare no competing interests.

\bibliography{references}

\clearpage

\setcounter{figure}{0}
\renewcommand{\figurename}{\textbf{Extended Data Fig.}}

\end{document}